\documentclass[12pt]{iopart}

\usepackage{iopams}
\usepackage{graphicx}
\usepackage{dcolumn}
\usepackage{bm}
\usepackage{color}

\newcommand{\fracc}[2]{\frac{\textstyle{#1}}{\textstyle{#2}}}

\begin{document}

\title{Radiating Spherical Collapse for an Inhomogeneous Interior Solution}

\author{Eduardo Bittencourt$^1$, Vanessa P Freitas$^{2}$, Jos\'e M Salim$^2$, Grasiele B Santos$^{1,2}$}
\address{$^1$ Universidade Federal de Itajub\'a, Av. BPS, 1303, Itajub\'a-MG, 37500-903, Brazil}
\address{$^2$ Centro Brasileiro de Pesquisas F\'isicas, Rua Doutor Xavier Sigaud 150, Rio de Janeiro, Brazil}
\eads{\mailto{bittencourt@unifei.edu.br},  \mailto{vpachecof@gmail.com}, \mailto{jsalim@cbpf.br}, \mailto{gbsantos@unifei.edu.br}}
\date{\today}

\begin{abstract}
We analyze the problem of gravitational collapse considering the matching of an exterior region described by the Vaidya's metric and an interior region described by a spherically symmetric shear-free inhomogeneous geometry sourced by a viscous fluid. We establish initial and final conditions for the process in order that the outcome be a non-singular object, when this is possible, and check how it depends on the fulfillment of the energy conditions. We then apply explicitly the matching procedure to the cases of linear and nonlinear Lagrangians describing electromagnetic fields inside the star, and analyze how the different behaviors for the scale factor of the interior geometry produce singular or nonsingular final stages of the collapse depending on the range where the initial conditions lie.

\end{abstract}

\maketitle

\section{Introduction}\label{I}
The process of relativistic gravitational collapse is of fundamental importance in astrophysics in order to understand how the formation of compact stars and black holes are possible, usually following a period where the star loses energy through radiation emission. The method used to describe such radiating collapse treats the star surface as dividing the spacetime in two disjoint regions: one interior, filled in with matter and radiation, and one exterior composed of all the radiation emitted by the star. These two submanifolds must be matched smoothly through a timelike hypersurface representing the evolving surface of the star in order to guarantee that the resulting spacetime is also a solution of Einstein's field equations. Spherical symmetry is often argued as a good approximation to astrophysical objects, rendering Vaidya's geometry \cite{vaidya51}, which is the unique spherically symmetric one purely sourced by radiation, as the appropriate metric to describe the exterior region. For the interior region, it has been shown \cite{fayos92} that every spherically symmetric metric can be matched to a Vaidya radiating solution provided that the total radial pressure vanishes on the star's surface. In particular, non-perfect interior fluids are considered in order to account for dissipation, heat diffusion and streaming processes that should be present in a radiating star \cite{bonnor89, fayos91}.

Once the interior and exterior geometries have been set, appropriate matching conditions should be imposed in order to glue such solutions in a general relativistic framework. Intrinsic Darmois' conditions \cite{darmois} have often been used to accomplish this task since it is a framework that does not require the same coordinatization in both regions \cite{bonnor81}. In this method, it is imposed the continuity of the first and second fundamental forms on the star's surface. The resulting equations, if consistent, then relate quantities of the exterior with quantities defined on the interior and give the equation describing how the star's surface evolves in time. In the analysis of gravitational collapse, different final stages of evolution have been found concerning the formation of singularities in general relativity: the singular ones, as the usual black hole solutions \cite{wald,joshi}, and the non-singular ones, including stable equilibrium configurations and bouncing solutions (usually associated with exotic fluids inside the star, see for example \cite{mbonye05,irina03} and references therein), or even the complete evaporation of the star \cite{fayos96,fayos08}.

In this work we focus on a particular shear-free interior solution with a non-trivial topology, that asymptotically behaves as the Friedmann-Lema\^itre-Robertson-Walker (FLRW) geometry. Its spatial section is the same as the Schwarzchild-de Sitter geometry, depending on the spatial curvature and an arbitrary parameter $M$ that renders the solution inhomogeneous. Following the same idea described in \cite{bitt14,dyn_worm}, we shall consider here both linear and nonlinear electromagnetic fields as sources for this geometry. For the sake of completeness, we start the next section by reviewing the general features of this solution as the interior geometry and analyze its energy-momentum content. Then, in Section \ref{III}, we briefly present the Vaidya solution that will be used as the exterior geometry describing, in Sec.\ \ref{IV}, how the solutions are glued together and how the collapse evolves. In Sec.\ \ref{V} we establish what should be the physical requirements for the initial and final stages of the star in order to have the formation of a stable object in general. Afterwards, in Sec.\ \ref{VI}, we analyze these conditions both in the case of an interior solution with a singular scale factor and of an interior solution with a bouncing scale factor, according to the degree of linearity of the electromagnetic field. Finally, we present some concluding remarks on the result of such radiating gravitational collapse. We set the units such that $c=\kappa=1$, where $c$ is the speed of light and $\kappa$ is the Einstein constant.

\section{Interior solution} \label{II}
In most of the circumstances in which a spherical symmetric gravitational collapse is manageable, the metric associated to the interior solution can be put in the form
\begin{equation}
\label{fried}
ds^2=-dt^2+a^2(t)[d\chi^2+\sigma^2(\chi)d\Omega^2],
\end{equation}
where $a(t)$ is the scale factor and $\sigma(\chi)$ is an arbitrary function of the spatial coordinate $\chi$, both of which are completely determined via Einstein's equations for a given matter content satisfying the symmetries of this metric.

In a first moment, we  shall consider solely that the Ricci scalar $^{(3)}R$ of spatial sections at constant $t$ of the metric (\ref{fried}) are constant and that this metric is compatible with an energy-momentum tensor such that the radial component of the diagonal pressure tensor can be different from those components along the angular directions. Under these assumptions, the spatial sector of (\ref{fried}) admits a unique solution, apart from the choice of parameters (see details in \cite{bitt14}) where $\sigma(\chi)$ must satisfy
\begin{equation}
\label{eq_sigma}
\sigma_{\chi}=\pm\sqrt{1-\frac{2M}{\sigma}-k\sigma^2},
\end{equation}
with $M$ being a constant of integration and $k$ being the constant spatial curvature with possible values $+1, 0,-1$. Making the coordinate transformation $r=\sigma(\chi)$, the line element given by (\ref{fried}) becomes
\begin{equation}
\label{fried_r_pi}
ds^2=-dt^2+a^2(t)\left(\fracc{dr^2}{b^2(r)}+r^2d\theta^2+r^2\sin^2{\theta}d\phi\right).
\end{equation}
where $b^2(r)=1-k r^2-2M/r$. We are restricted to the case where $M$ is non-negative, otherwise the metric tensor would possess a naked singularity at $r=0$. Thus, there is a minimum value for $r$ for a given $k$. The physical meaning of the parameter $M$ is related to the scale of homogeneity of the solution: whenever $2M/r\ll1$, this term can be neglected in (\ref{fried_r_pi}) and, hence, the FLRW geometry is recovered. On the other hand, if $2M/r$ is not negligible, the spatial part of that geometry is the same as the Schwarzschild-de Sitter metric.

With this in mind, a possible source for this solution is a non-perfect fluid represented by an energy-momentum tensor of the form
\begin{equation}
\label{tmunu}
T_{\mu\nu} = \rho v_\mu v_\nu +p_n n_{\mu}n_{\nu}+p_t(\omega_{\mu}^{\theta}\omega^{\theta}_{\nu}+\omega_{\mu}^{\phi}\omega^{\phi}_{\nu}) ,
\end{equation}
where $\{v_\mu,n_\mu,\omega_{\mu}^{\theta},\omega_{\mu}^{\phi}\}$ is an orthonormal basis such that $\omega^\theta=a(t)rd\theta$, $\omega^\phi=a(t)r\sin\theta d\phi$ and $v^\mu$ is the four-velocity of the fluid. For an observer comoving with the fluid, $\rho$ is the energy density, $p_n$ is the normal pressure and $p_t$ is the tangential pressure. This source is equivalent to the one presented in \cite{bitt14} with the components of the diagonal anisotropic pressure tensor $\pi^{\mu}{}_{\nu}$ given by
\begin{equation}
\label{p_munu_func}
\pi^1{}_1=\frac{2}{3}(p_n-p_t),\quad\mbox{and}\quad \pi^2{}_2=-\frac{1}{3}(p_n-p_t)=\pi^3{}_3,
\end{equation}
and the isotropic pressure being the average over the directions, that is
\begin{equation}
\label{p_func}
p=\frac{p_n+2p_t}{3}.
\end{equation}
It is important to note that this solution is shear-free and, thus, the anisotropic pressure must be linked to some other geometrical quantity, which in our case is the electric part of the Weyl tensor.

For the above described geometry, the Einstein's field equations $G_{\mu\nu}= T_{\mu\nu}$ lead to the following system
\begin{eqnarray}
&&\frac{a_{,t}^2}{a^2}+\frac{k}{a^2}=\frac{1}{3}\rho,\label{0}\\
&&2\frac{a_{,tt}}{a}+\frac{(a_{,t}^2+k)}{a^2}+\frac{2M}{a^2r^3}=-p_n,\label{1}\\
&&2\frac{a_{,tt}}{a}+\frac{(a_{,t}^2+k)}{a^2}-\frac{M}{a^2r^3}=-p_t,\label{2}
\end{eqnarray}
where we denote $a_{,t}\equiv da/dt$ and $a_{,tt}\equiv d^{\,2}a/dt^2$. We omit a possible contribution from the cosmological constant to the equations above, but it can be recovered at any moment through the map $\rho\rightarrow\rho+\Lambda$ and $p_{(t,n)}\rightarrow p_{(t,n)}-\Lambda$. In terms of isotropic and anisotropic pressures it is straightforward to see, using (\ref{p_munu_func}) and (\ref{p_func}), that Einstein's equations demand $p=p(t)$ and the components of $\pi^{\mu}{}_{\nu}$ to be given by
\begin{equation}
\label{pi_munu}
\pi^1{}_1=-\frac{2M}{a^2r^3},\quad\mbox{and}\quad \pi^2{}_2=\pi^3{}_3=-\frac{\pi^1{}_1}{2}.
\end{equation}
Note that these results do not constrain the evolution of the scale factor.

Another useful quantity in the analysis of the loss of mass by the star through radiation emission is the mass function, whose definition is given in \cite{hern66,cahill70}, and which reads for this solution
\begin{equation}
\label{mass}
m(t,r)=\frac{a r}{2}\, R^3{}_{232}=\frac{a}{2}\left(r^3a_{,t}^2+k r^3+2M \right).
\end{equation}
If the scale factor is not constant in time, we can use (\ref{0}) to rewrite the mass function in a more suitable way as
\begin{equation}
\label{mass_func}
m(r,t)=aM+\frac{r^3a^3}{6}\,\rho.
\end{equation}
Whenever $a\neq0$ and $M\neq0$, the mass function never goes to zero during the collapse and also $r=0$ is never achieved due to the existence of a minimum value for the proper radius of the star. This fact would suggest the existence of a non-singular object as a product of the collapse which, however, will happen only for very specific initial conditions, as we shall see in what follows.

\section{Exterior solution}\label{III}
We consider a radiating exterior solution described by Vaidya's metric \cite{vaidya51} written in radiative coordinates \cite{bondi62} as
\begin{equation}
ds^2=-\bar{\chi}\,du^2+2\varepsilon\, dudR+R^2(d\bar{\theta}^2+\sin^2\bar{\theta}d\bar{\phi}^2),
\end{equation}
where $\varepsilon^2=1$, $\bar{\chi}=1-2\bar{m}(u)/R$ and the mass function seen from the exterior $\bar{m}(u)$ depends only on $u$. The source for this geometry is considered to be pure incoherent radiation represented by the following energy-momentum tensor
\begin{equation}
\label{Text}
\bar{T}_{\mu\nu}=\frac{2\varepsilon}{R^2}\frac{d\bar{m}(u)}{du}l_\mu l_\nu,
\end{equation}
where $l_\mu$ is a null vector satisfying $l_\mu dx^\mu=-du$. Clearly the weak energy condition implies that $\varepsilon\frac{d\bar{m}(u)}{du}\geq 0$. Assuming that $u$ grows towards the future, we have two possibilities: either $\bar{m}$ is a non-increasing function of $u$ for $\varepsilon=-1$ and the radially directed radiation described by the above energy-momentum tensor is outgoing (increasing values of the coordinate $R$) or $\bar{m}$ is a non-decreasing function for $\varepsilon=1$ and the radiation is ingoing (decreasing values of the coordinate $R$). The choice will be made after the matching between the exterior and the interior solutions when we set the direction of the radiation emitted by the star.

\section{The matching procedure} \label{IV}
In order to have a complete stellar solution, we should be able to match the interior and exterior spacetimes through the evolving star's surface which is a timelike hypersurface that will be denoted $\Sigma$. According to Darmois junction conditions, the first and second fundamental forms of the interior and exterior geometries should be continuous through such hypersurface. Using indexes $+$ and $-$ to denote the exterior and the interior, respectively, we first consider a general timelike hypersurface $\Sigma_{+}$ that preserves the spherical symmetry of Vaidya's solution with intrinsic coordinates $\{\xi^a\}=\{\tau,\vartheta,\varphi\}$ and defined by the parametric equations $u=u(\tau)$, $R=R(\tau)$, $\bar{\theta}=\vartheta$ and $\bar{\phi}=\varphi$. In an analogous manner\footnote{The intrinsic coordinates are chosen to be the same in both hypersurfaces $\Sigma_{+}$ and $\Sigma_{-}$ as we assume a diffeomorphism between them. For details see \cite{fayos96}.}, we take a timelike hypersurface $\Sigma_{-}$ that preserves the spherical symmetry of the solution (\ref{fried_r_pi}) defined by $t=t(\tau)$, $r=r(\tau)$, $\theta=\vartheta$ and $\phi=\varphi$.

There are two embeddings of $\Sigma$, one defined by $\Sigma_{+}=\Sigma$ and the other given by $\Sigma_{-}=\Sigma$. Each of them induces a metric on $\Sigma$ via the formulae
\begin{equation}
\label{first_fund}
g^{\pm}_{ab}=g^{\pm}_{\mu\nu}\frac{\partial x^{\mu}_{\pm}}{\partial \xi^{a}}\frac{\partial x^{\nu}_{\pm}}{\partial \xi^{b}}.
\end{equation}
In its turn, from each metric we can construct a line element on $\Sigma$ as $ds_{\pm}^2=g^{\pm}_{ab}\,d\xi^ad\xi^b$. Therefore, when $\Sigma_{+}=\Sigma=\Sigma_{-}$, the first Darmois condition requires that
\begin{equation}
\label{mat_1_form}
ds_+^2=ds_-^2.
\end{equation}

The components of the second fundamental form are given by the extrinsic curvature of $\Sigma_{\pm}$, which reads
\begin{equation}
K^\pm_{ab}=-n^\pm_\alpha\frac{\partial}{\partial \xi^a}e_{b}^{\  \alpha} - n^\pm_\alpha\, (\Gamma^\alpha_{\mu\nu})^{\pm} e_{a}^{\  \mu}e_{b}^{\ \nu},
\end{equation}
where $n^{\pm}_{\mu}$ are unit normal vectors, $(\Gamma^\alpha_{\mu\nu})^{\pm}$ are the Christoffel symbols and $e_{a}^{\pm \mu}={\partial x^\mu_{\pm}}/{\partial \xi^a}$ on both sides of the boundary surface $\Sigma$. The set $\{n^{\mu},e_{a}^{\mu}\}$ forms a basis of the tangent space on $\Sigma$. Thus, the second Darmois condition imposes that
\begin{equation}
\label{mat_2_form}
K_{ab}^+=K_{ab}^-.
\end{equation}

In this case, the unit normal vectors read
\begin{equation}
n_\mu^-=\frac{\epsilon a \dot{t}}{\sqrt{b^2\dot{t}^2-a^2\dot{r}^2}}\left(-\frac{\dot{r}}{\dot{t}},1,0,0\right)
\end{equation}
and
\begin{equation}
n_\mu^+=\frac{\bar{\epsilon}}{\sqrt{\dot{u}(\bar{\chi}\dot{u}-2\varepsilon\dot{R})}}\left(-\dot{R},\dot{u},0,0\right),
\end{equation}
where dot means derivative with respect to $\tau$. The quantities $\epsilon$ and $\bar{\epsilon}$ give the orientation of these vectors and they are such that $\epsilon^2=1$ and $\bar{\epsilon}^2=1$. We shall find the appropriate signs after obtaining the equations that describe the matching.

The matching conditions applied to the first fundamental form (\ref{mat_1_form}) lead to the following equations
\begin{eqnarray}
\label{h00}&&\dot t^2-\frac{a^2\dot r^2}{b^2}=\bar\chi\dot u^2-2\varepsilon\dot u\dot R,\\[2ex]
\label{h22}&&R=ar.
\end{eqnarray}
We see that (\ref{h00}) associates the first derivatives of the time and radial coordinates on both sides of $\Sigma$, while (\ref{h22}) relates the radii of the 2-spheres in each coordinate system.

After some manipulations, the continuity of the second fundamental form (\ref{mat_2_form}) across $\Sigma$ yields
\begin{eqnarray}
\label{udot}&&\dot u=\frac{1}{\bar\chi}\left[\epsilon\bar\epsilon\left(\frac{r\dot raa_t}{b}+b\dot t\right)+\varepsilon\dot R\right],\\[2ex]
\label{rdot}&&\dot r=\epsilon\bar\epsilon\varepsilon\, \frac{b}{a}\frac{p_n}{\rho}\dot t.
\end{eqnarray}

From (\ref{h00}) and (\ref{udot}) we can write
\begin{equation}
\label{chi}
\bar\chi=b^2-r^2a_{,t}^2,
\end{equation}
and this relation implies that the mass function of the Vaidya metric can be written only in terms of quantities associated to the interior solution as
\begin{equation}
\label{massfunction}
\bar m(t,r)=\frac{a\,r^3(a_{,t}^2+k)}{2}+aM,
\end{equation}
which matches (\ref{mass}) precisely, as we expected. The behaviour of the mass function in terms of $\tau$ can be better understood through its time derivative. Therefore, substitution of (\ref{rdot}) into the time derivative of (\ref{massfunction}) leads to
\begin{equation}
\label{der_mas_func}
\dot{\bar m}=\frac{r^2a^2}{2}(\epsilon\bar\epsilon\varepsilon b-r a_t)p_n.
\end{equation}

In order to have radiation coming out of the star during the collapse we should set $\varepsilon=-1$, so that the radially directed radiation described by (\ref{Text}) be outgoing. Besides, by substituting (\ref{rdot}) in (\ref{udot}) and using (\ref{chi}), we get
\begin{equation}
\frac{\dot{u}}{\dot{t}}=\left(\frac{\rho+p_n}{\rho}\right)\frac{1}{\epsilon\bar{\epsilon}b+ra_{,t}}.
\end{equation}
This ratio is bigger than zero, as we consider that both $u$ and $t$ grow toward the future (increasing values of $\tau$). Thus, if the null energy condition is satisfied ($\rho+p_n>0$) and $a_{,t}<0$ (due to the collapse), then we should have $\epsilon\bar{\epsilon}=+1$. With this we set the appropriate signs appearing in the matching equations. Note that all variables corresponding to the exterior solution are determined by the ones of the interior. From now on, we choose the parametrization $t(\tau)=\tau$ which is the only arbitrariness in the equations above. Thus, the time derivatives are now with respect to the time coordinate $t$, in particular, the dynamics of the star surface given by (\ref{rdot}), which is the only equation we need to solve, since all the other variables can be determined by quadrature for a given matter content.

\section{Boundary conditions for a collapsing star}\label{V}
The set of equations derived above together with the Friedmann equation (\ref{0}) and the continuity equation define an autonomous dynamical system when an equation of state $p=p(\rho)$ is provided. Such system has no equilibrium points if the dominant energy condition for the interior holds throughout the collapse. This means that the star has no preferred initial or final configurations, mathematically speaking. Therefore, we should select from all the possibilities solely those which have a reasonable physical meaning. In this way, the gravitational collapse will be characterized here by initial conditions such that the proper radius of the star and its total mass decrease. In other words, we should have
\begin{equation}
\label{in_conf}
r_{,t}(t=t_0)<0,\quad \mbox{and}\quad m_{,t}(t=t_0)<0,
\end{equation}
where $t_0$ corresponds to the moment in time in which the star starts to collapse. These assumptions imply that the radial pressure $p_n$ must be positive when the star begins to contract, while the time derivative of the scale factor must be on the interval $0<-a_{,t}<b/r$. If the system reaches the upper limit along the collapse, i.e., $-a_{,t}(t_c)=b(r(t_c))/r(t_c)$ at a given instant of time $t_c>t_0$, then the radius of the star crosses the apparent horizon, defined by $\bar\chi=0$ in (\ref{chi}). Thus, $\Sigma$ becomes a trapped surface and the collapse will eventually end up in a singularity.

With the help of (\ref{p_munu_func}), (\ref{p_func}) and (\ref{pi_munu}), the positivity of $p_n$ establishes a lower bound for the initial radius of the star given by
\begin{equation}
\label{cond_r0_1}
r_0>\left(\frac{2M}{p_0a_0^2}\right)^{\frac13},
\end{equation}
where $p_0=p(t_0)$ is the initial value of the barotropic pressure and $a_0=a(t_0)$ is the initial value of the scale factor. Note that this inequality is general enough to hold for any equation of state. On the other hand, the fact that the initial spherical surface of the star must be anti-trapped requires that $\bar{\chi}>0$, that is
\begin{equation}
\label{cond_r0_2}
\left(1-\frac{2M}{r}\right)\frac{1}{r^2}>\frac{a_0^2\rho_0}{3},
\end{equation}
where $\rho_0$ is the initial energy density. The left-hand side of this inequality imposes that $r>2M$ (independently of the spatial curvature), having a maximum at $r=3M$ and going asymptotically to zero as $r$ goes to infinity. Therefore, once $a_0$, $\rho_0$ and $p_0$ are fixed, (\ref{cond_r0_1}) and (\ref{cond_r0_2}) should be combined in order to get the range of possible initial values for the radius of the star.

The final stages of the collapse can be formulated in terms of the fulfillment of the null energy condition at the exterior. Since $m_{,t}$ is a decreasing function on the interior and the rate of radiation emitted by the star is directly linked to it, when it vanishes before crossing the apparent horizon, the radiation emission ceases and the energy momentum tensor at the exterior becomes null. Due to Birkhoff's theorem, the exterior solution then becomes the Schwarzschild metric if the total mass is nonzero. In such phase, the proper radius of the star continues to diminish due to the evolution of the scale factor, although the exterior geometry is static \cite{adler05}. If there is no mechanism to stop the contraction of the scale factor, the whole dynamics will indeed lead to a singularity. Therefore, in order to have a non-singular gravitational collapse, $a_{,t}$ must remain in the interval mentioned above for all time and the time derivative of the mass function should vanish before the system crosses the apparent horizon, demanding $p_n=0$ and, consequently, $r_{,t}$=0. Indeed, from (\ref{massfunction}), we see that this is the only possibility of a non-singular final stage, since the mass function vanishes only if the scale factor goes to zero and the energy conditions are satisfied. It should also be noticed that the complete evaporation of the star leading to the Minkowski spacetime is possible only for $M=0$, which is the case described in \cite{fayos08}.

\section{Consequences of the scale factor evolution for the collapse}\label{VI}
Now, we shall study the gravitational collapse by making some considerations about the dynamics of the scale factor and our goal is to find physical conditions in which the collapse is a regular process. In particular, we will divide the discussion in two cases: singular and  nonsingular scale factors. In the former, we analyze the role played by the constant $M$ in changing the possible final stages of the star. In the latter, since some of the energy conditions must be violated, we  seek for sufficient conditions in order to have a physical reasonable evolution for the star. Recall that in virtue of the attractive nature of gravity, a purely gravitational collapse will always end up into a singularity if the content of the star is a fluid satisfying the energy conditions, unless the star evaporates. Furthermore, it is not obvious to set a suitable equation of state such that $m_{,t}\rightarrow0$ before crossing the apparent horizon. In general, other mechanisms related to thermodynamics or kinetic theory should be taken into account to prevent the collapse to a singularity also rendering the model more realistic.

\subsection{Singular scale factor}
Models aiming at a regular gravitational collapse in which the interior solution has a contracting singular scale factor require that the star achieves some stable configuration before the singularity is reached. However, if the energy conditions are satisfied, the final stages will eventually lead to the formation of a black hole (see \cite{joshi,port} and references therein). The only possibility to avoid the singularity satisfying the energy conditions is through the evaporation of the star as discussed in \cite{fayos08}. In that reference, the authors study a class of models in which the star radiates away all of its matter content before crossing the apparent horizon. For the sake of comparison, their models correspond to $M=0$ in our approach and, therefore, it is not necessary to discuss this case to avoid redundancy.

Let us analyze the case when $M\neq0$. In this situation, the evaporation is not possible since the mass function has now a contribution of the form $aM$ which does not vanish unless $a$ goes to zero [see (\ref{mass_func})]. Besides, if $p_n\geq0$ along the collapse, the mass function must satisfy the following inequality
\begin{equation}
m(t,r)\leq\frac{(ar)^3}{6}\,(\rho+3p),
\end{equation}
which means that it cannot be zero, for $a\neq 0$, unless $\rho+3p$ vanishes. Since this is not the case here (such a possibility with violation of the strong energy condition will be treated in the next section), then we conclude that any collapse process where the interior solution is modeled by the metric (\ref{fried_r_pi}) with $M\neq0$ and a singular scale factor cannot evaporate and will always end up into a singularity.

\subsection{Nonsingular scale factor}
It is not possible to study nonsingular contracting models for the scale factor without discussing the energy conditions. In particular, it is well known that there should be violation of the strong energy condition in order to have a bounce in the scale factor \cite{nov_sepb}. Therefore, if one desires to see what happens to the gravitational collapse when the scale factor is nonsingular, the whole system of equations obtained from the junction conditions should be recast. Up to this point, we have assumed that the energy conditions all hold and we have tried to follow the path towards a nonsingular gravitational collapse. Under these assumptions, we set the initial and final stages for the collapse as mentioned before and, besides, the directions of the unit normal vectors through the signs of $\epsilon$ and $\bar\epsilon$.

Notwithstanding, there are many ways to approach this issue and our aim is not to scrutinize all possibilities. In what follows, we shall focus on the study of a particular model setting initial conditions for the star as stated before, assuming a nonlinear Lagrangian as source for the gravitational field within the star. We will not seek for an specific final stage {\it a priori}, but we will let the equations  indicate the more appropriate possibilities of final configuration for the star.

Let us consider a nonlinear Lagrangian of the form
\begin{equation}
\label{non_lag}
L=-\frac{F}{4}+\alpha F^2,
\end{equation}
where $F=\,F^{\mu\nu} F_{\mu\nu}$, with $F_{\mu\nu}$ as the Faraday tensor and the arbitrary parameter $\alpha$ is positive in order to have a nonsingular scale factor \cite{deLorenci}. In virtue of the high electric conductivity inside the star, we can assume as a simplification that only the magnetic field $B(t)$ contributes, in average, to the interior solution. In this case, the energy density and the isotropic pressure are given by
\begin{equation}
\rho=\frac{B^2}{2}(1-8\alpha B^2),\quad\mbox{and}\quad p=\frac{B^2}{6}(1-40\alpha B^2),
\end{equation}
while the time evolution of the magnetic field itself is written as
\begin{equation}
B(t)=\frac{B_0}{a^2(t)}.
\label{B_a}
\end{equation}
When the spatial curvature vanishes ($k=0$), we can use (\ref{0}) to find an explicit expression for the scale factor in terms of t, as follows
\begin{equation}
\label{sol_a_flat}
a(t)=a_b\left[\left(\frac{t}{t_c}\right)^2+1\right]^{\frac{1}{4}},
\end{equation}
where $a_b=(8\alpha B_0^2)^{\frac{1}{4}}$ is the size of the scale factor at the bounce and $t_c=\sqrt{12\alpha}$ is the instant of time in which the energy density reaches its maximum value.

\begin{figure}
\begin{center}
\includegraphics[width=45mm]{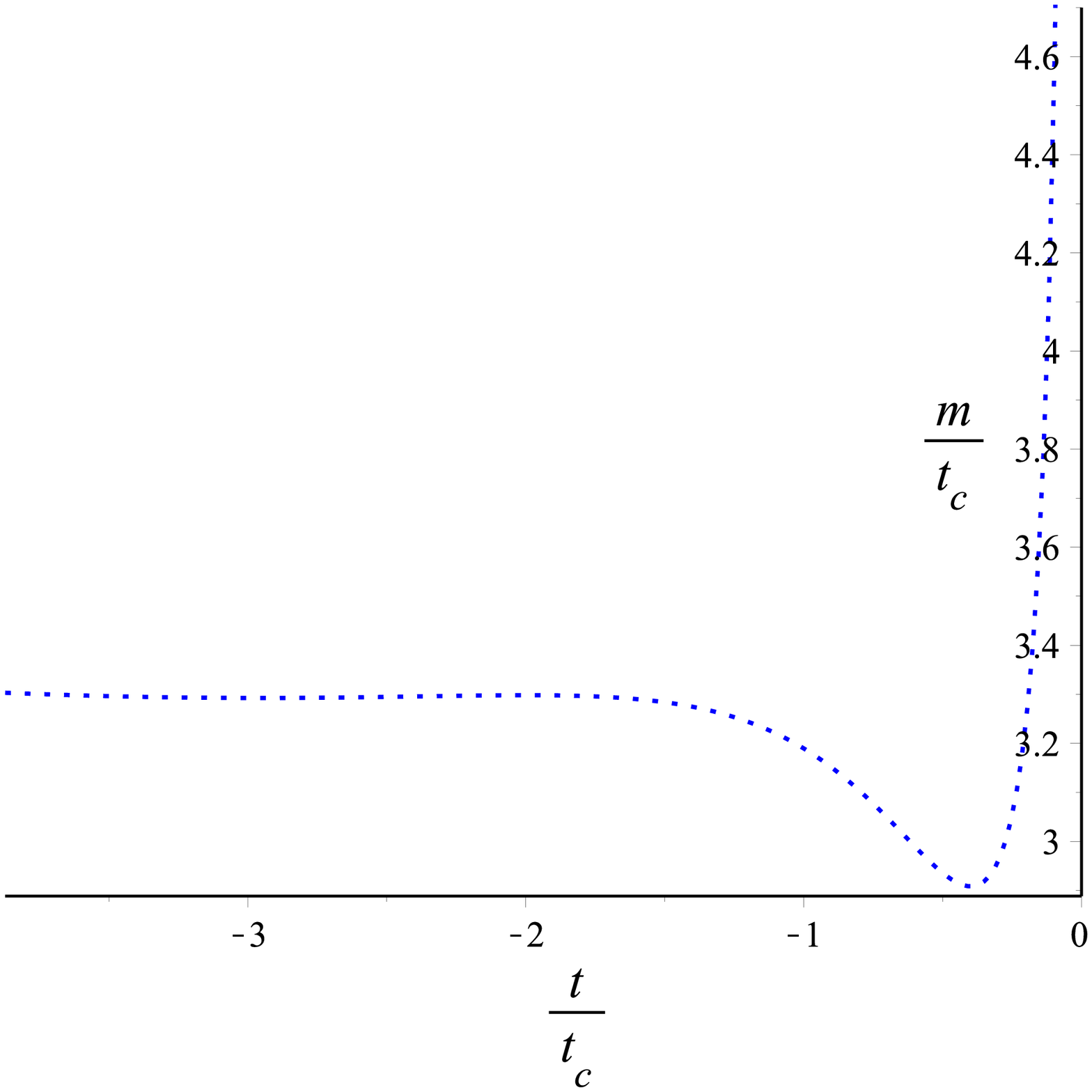}\hspace{.5cm}
\includegraphics[width=45mm]{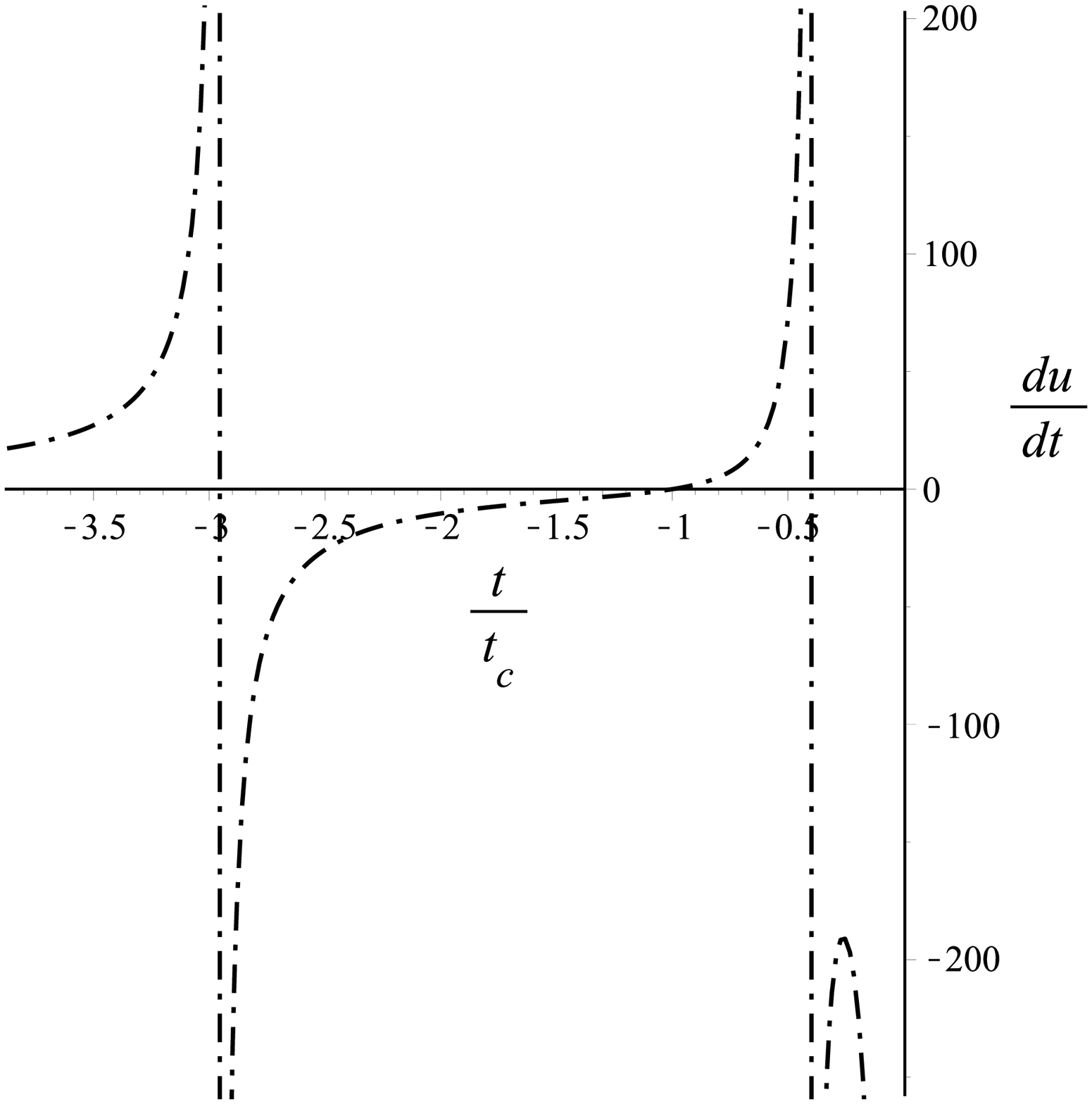}\hspace{.5cm}
\includegraphics[width=45mm]{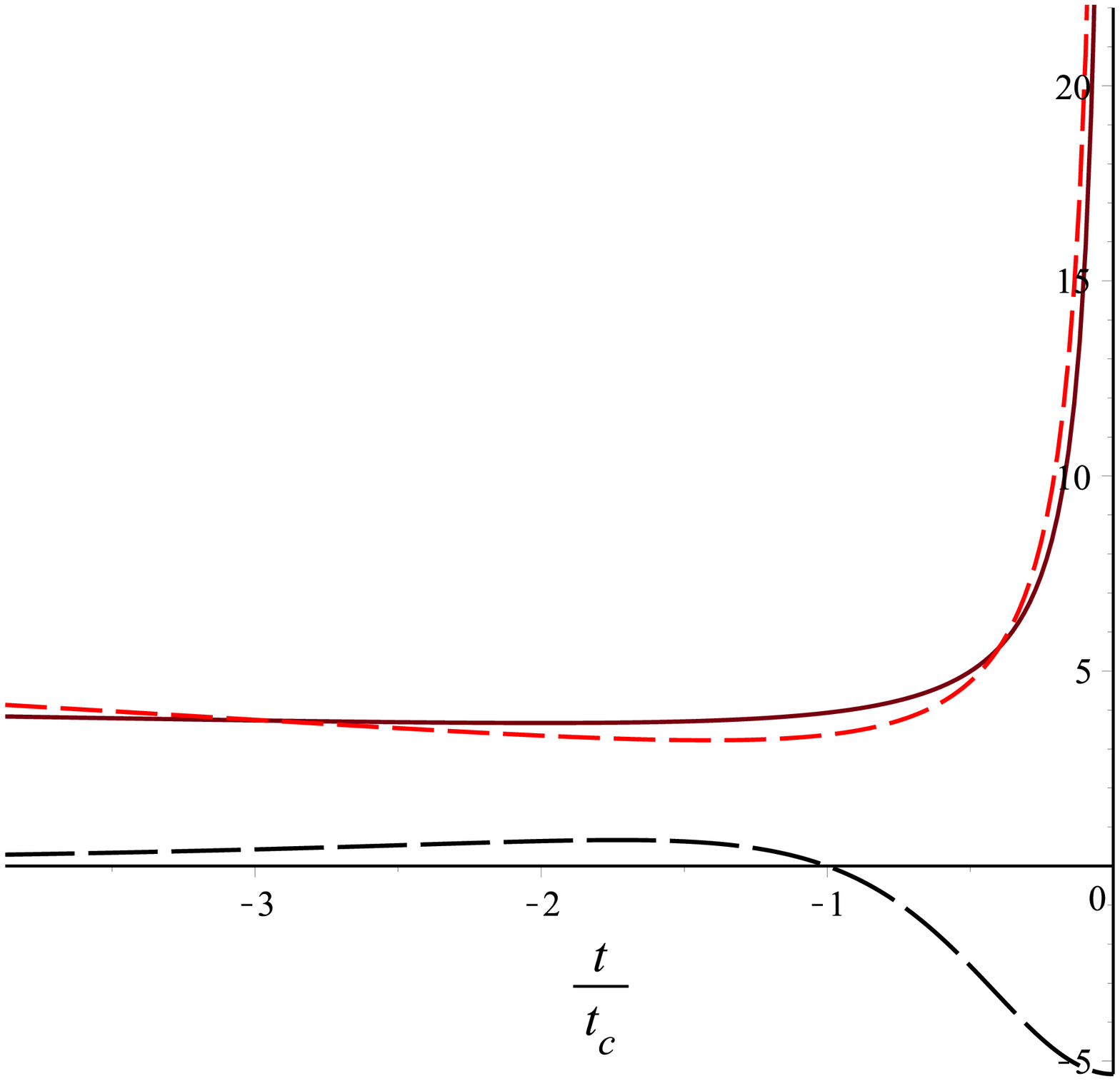}
\end{center}
\caption{Case $M=0$. Illustrative initial values are used as $\tilde{t}_0=-\sqrt{15}$ and $\tilde {r}_0=\tilde {r}(\tilde{t}_0)\approx3.8$. On the left panel, the dotted line represents $\tilde{m}(\tilde t)$. On the middle panel, the dot-dashed line corresponds to $d\tilde u/d\tilde t$. On the right, it is depicted $\tilde r(\tilde t)$ (solid line), $\tilde r_{A3H}(\tilde t)$ (dashed line) and the quantity $\rho+p_n$ (long-dashed line). $\tilde {r}_0$ was chosen to be initially a bit smaller than the apparent 3-horizon at $t=t_0$.}
\label{fig1}
\end{figure}

In a first moment, we study the collapse considering $M=0$ in (\ref{fried_r_pi}). We then define new dimensionless variables $\tilde t=t/t_c$, $\tilde r=r\,a_b/t_c$, $\tilde u=u/t_c$ and $\tilde m=m/t_c$ and rewrite all the equations of interest in terms of them, obtaining
\begin{eqnarray}
&&\frac{d\tilde r}{d\tilde t}=-\frac{\tilde{t}^{\,2}-4}{3\,\tilde{t}^2(\tilde{t}^{\,2}+1)^{\frac{1}{4}}},\label{rdot_b}\\[1ex]
&&\frac{d\tilde u}{d\tilde t}=\left(1 + \frac{\tilde{t}^{\,2} - 4}{3\,\tilde t^{\,2}}\right) \left[1 - \frac{\tilde r\, |\tilde t|}{2\,(\tilde t^{\,2}+1)^{\frac{3}{4}}} \right]^{-1}, \label{udot_b}\\[1ex]
&&\tilde m=\frac{\tilde{r}^3\tilde{t}^{\,2}}{8(\tilde{t}^{\,2}+1)^{\frac{5}{4}}},\label{m_fun_b}\\[1ex]
&&\tilde{r}_{A3H}=\frac{2\,(\tilde{t}^{\,2}+1)^{\frac{3}{4}}}{|\tilde t|}.\label{r_a3h_b}
\end{eqnarray}
We remark that (\ref{rdot_b}) can be fully integrated and the outcome is given in terms of hypergeometric functions, which will be useful for plotting the curves $\tilde r(\tilde t)$ but are otherwise too clumsy to be written here explicitly. Also, from (\ref{rdot_b}), one sees that an initial condition with $d\tilde r/d\tilde t<0$ is possible only if $\tilde t<-2$. Furthermore, the initial radius of the star should be smaller than the apparent 3-horizon (\ref{r_a3h_b}) in order that $\Sigma$ be untrapped. With this in mind, we depict the results in Fig.\ (\ref{fig1}).
From there we see that when $\tilde r(\tilde t)$ crosses $\tilde{r}_{A3H}$, $\dot u$ diverges and, as a consequence, an event horizon is formed. Therefore, the $u$-coordinate cannot be extended beyond its first divergence on the left part of the middle panel in Fig.\ (\ref{fig1}), and some analytic extension for the exterior solution is needed\footnote{We will not discuss the analytical continuation of Vaidya's metric in our case, since there is no consensus when energy conditions are violated (see \cite{berezin16} and references therein). However, the information about the star evolution provided only by the interior solution is enough for our purposes here.}. On the other hand, from the other panels at the same figure we notice that the interior solution is completely regular, except when $\tilde t$ goes to zero, where $\tilde r$, $\tilde{r}_{A3H}$ and $\tilde m$ diverge. Nevertheless, the ratio $\tilde m/\tilde r^3$, which is the dominant term in all curvature invariants, goes to zero in that limit, indicating that the curvature of the exterior solution vanishes there (see appendix in \cite{bitt14} for details).

In Fig.\ (\ref{fig2}), we select another initial value $\tilde{r}_0$ (smaller than the previous one) such that it does not cross the apparent 3-horizon along its whole evolution for negative values of time. However, when $\rho+p_n$ vanishes, $d\tilde u/d\tilde t$ becomes zero, indicating that the null energy condition is violated and that the $u$ coordinate cannot be used to describe the evolution on the exterior, demanding another time coordinate. Again, the other quantities behave well, but grow indefinitely as $t$ approaches zero. Yet, in this case, $\lim_{\tilde t\rightarrow0^{-}}(\tilde m/\tilde r^3)=0$ and the curvature becomes flat at $\tilde{t}=0$. In these first two examples mentioned above, the final stage of the star occurs at $\tilde t=0$ when it undergoes an infinite expansion in a finite time. A possible continuation of such solution for positive values of time could be performed by considering only the interior geometry, which will then coincide with the whole spacetime once $\tilde{r}(\tilde{t})\rightarrow\infty$.

\begin{figure}
\begin{center}
\includegraphics[width=70mm]{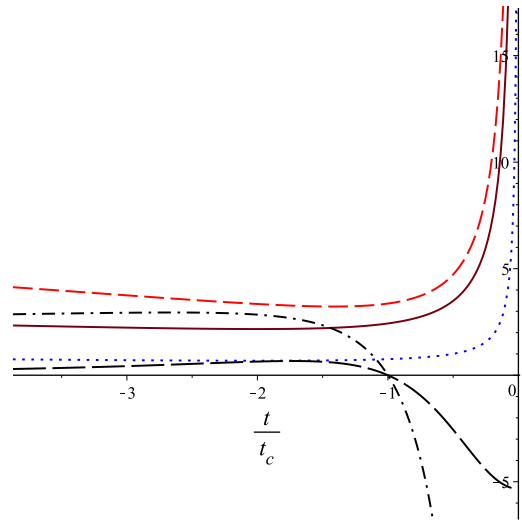}\hspace{.5cm}
\end{center}
\caption{Case $M=0$. Illustrative initial values are used as $\tilde{t}_0=-\sqrt{15}$ and $\tilde{r}_0\approx2.3$. It is depicted $d\tilde u/d\tilde t$ (dot-dashed line), $\tilde r(\tilde t)$ (solid line), $\tilde r_{A3H}(\tilde t)$ (dashed line), mass function (dotted line) and $\rho+p_n$ (long-dashed line).}
\label{fig2}
\end{figure}

\begin{figure}
\begin{center}
\includegraphics[width=60mm]{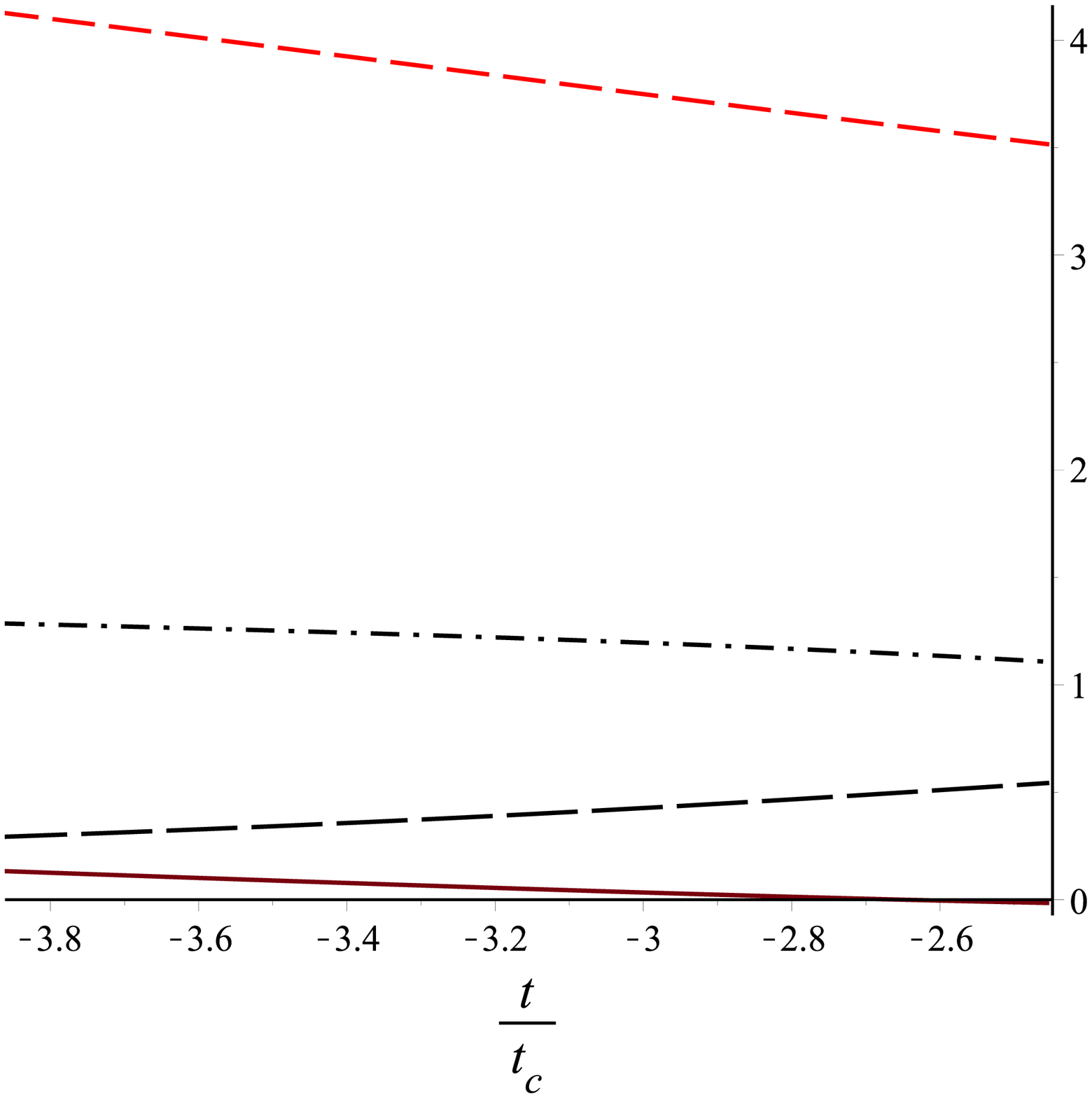}\hspace{.5cm}
\includegraphics[width=60mm]{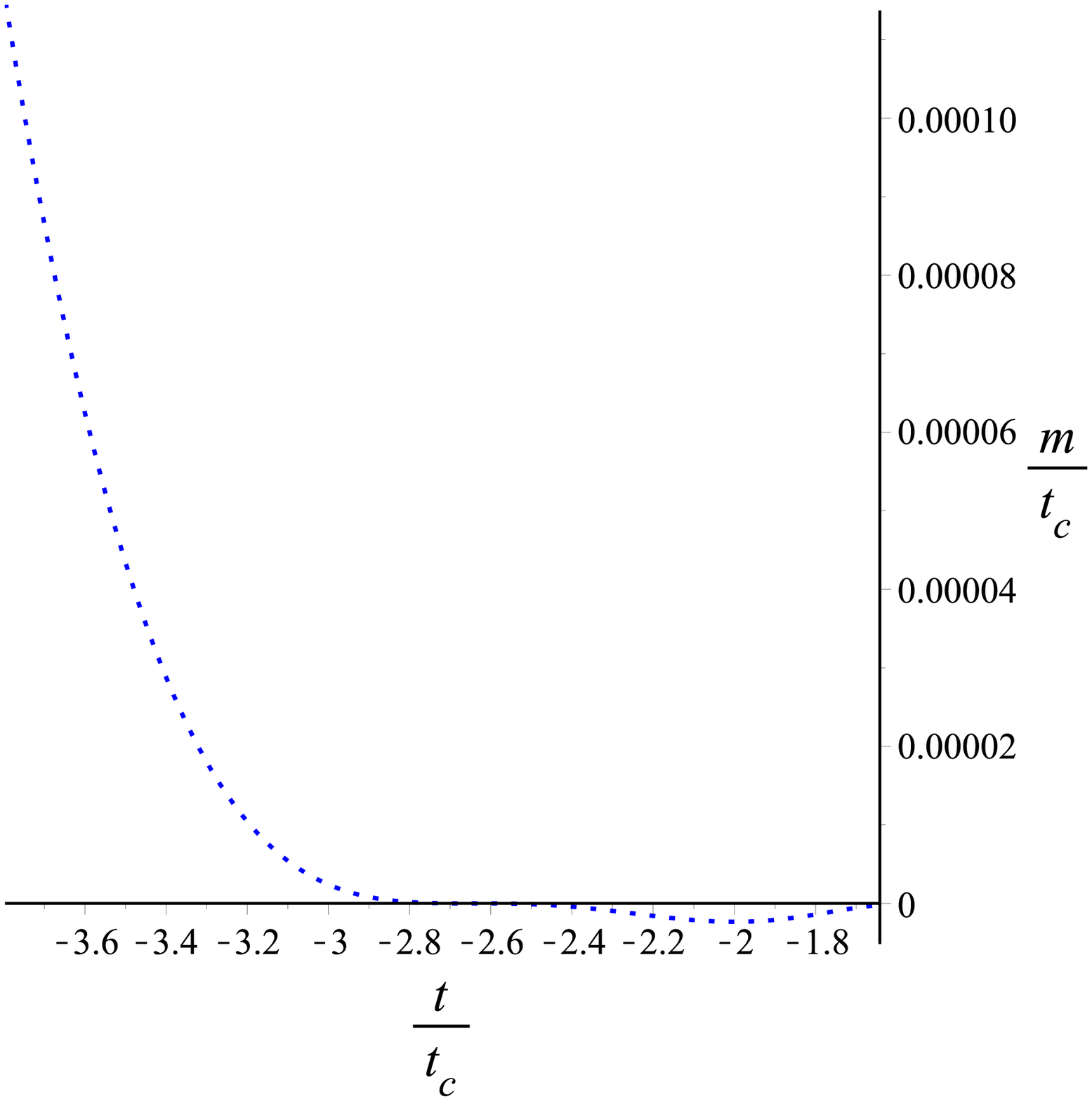}
\end{center}
\caption{Case $M=0$. Illustrative initial values are used as $\tilde{t}_0=-\sqrt{15}$ and $\tilde{r}_0\approx0.13$. On the left panel, it is depicted $d\tilde u/d\tilde t$ (dot-dashed line), $\tilde r(\tilde t)$ (solid line), $\tilde r_{A3H}(\tilde t)$ (dashed line) and $\rho+p_n$ (long-dashed line). On the right panel, the dotted line indicates the mass function. Note that $\tilde{r}_0$ and $\tilde{m}$ vanishes, simultaneously, around $\tilde{t}=-2.65$.}
\label{fig3}
\end{figure}

Finally, for sufficiently small values of $\tilde{r}_0$, it is possible to have the complete evaporation of the star at some negative value of time, as depicted in Fig.\ (\ref{fig3}). The window for such situation is quite narrow considering the set of initial conditions we used, but recall that all quantities are dimensionless and thus, in principle, can be re-scaled at our will.

For the case $M\neq0$ in (\ref{fried_r_pi}), we have tried to proceed similarly to the previous cases. Thus, we use the same set of dimensionless variables as before, but now $\tilde r$ is defined as $\tilde r=r/2M$ and a scale parameter denoted as $\beta=t_c/(a_bM)$ appears. The equations of interest are now written as
\begin{eqnarray}
\fl\qquad
&&\frac{d\tilde r}{d\tilde t}=-\frac{\beta\,\sqrt{1-\frac{1}{\tilde r}}}{6\tilde{t}^{\,2}(\tilde{t}^{\,2}+1)^{\frac{1}{4}}} \left[ \tilde{t}^{\,2} - 4 - \frac{\beta^2(\tilde{t}^{\,2}+1)^{\frac{3}{2}}}{\tilde r^{\,3}}\right],\label{rdot_b_m}\\[1ex]
\fl\qquad
&&\frac{d\tilde u}{d\tilde t}=\left\{1 + \frac{1}{3\,\tilde t^{\,2}}\left[\tilde{t}^{\,2} - 4 - \frac{\beta^2\,(\tilde t^{\,2}+1)^{\frac{3}{2}}}{\tilde r^{\,3}}\right]\right\} \left[\sqrt{1-\frac{1}{\tilde r}} - \frac{\tilde r\, |\tilde t|}{\beta\,(\tilde t^{\,2}+1)^{\frac{3}{4}}} \right]^{-1}, \label{udot_b_m}\\[1ex]
\fl\qquad
&&\tilde m=\frac{\tilde{r}^3\tilde{t}^{\,2}}{\beta^3\,(\tilde{t}^{\,2}+1)^{\frac{1}{4}}}+\frac{(\tilde{t}^{\,2}+1)^{\frac{1}{4}}}{\beta},\label{m_fun_b_m}
\end{eqnarray}
Now, the apparent 3-horizon curve is given by the solutions of the equation
\begin{equation}
\frac{\tilde{t}^{\,2}\,(\tilde{r}_{A3H})^3}{\beta^2\,(\tilde{t}^{\,2}+1)^{\frac{3}{2}}}-\tilde{r}_{A3H}+1=0,\label{r_a3h_b_m}
\end{equation}
which corresponds to a cubic equation for $\tilde{r}_{A3H}(\tilde t)$ and whose number of solutions depends of the sign of the discriminant
\begin{equation}
\Delta = \frac{\tilde{t}^{\,2}}{\beta^2\,(\tilde{t}^{\,2}+1)^{\frac{3}{2}}} \left(4 - 27 \frac{\tilde{t}^{\,2}}{\beta^2\, (\tilde{t}^{\,2}+1)^{\frac{3}{2}}}\right).
\end{equation}
If $\Delta=0$, we then have another cubic function, now in terms of $\tilde t$, which, in its turn, gives a critical value for the parameter $\beta_c=3^{\frac34}/\sqrt{2}$. Therefore, the allowed regions for the gravitational collapse initial conditions depend on the values of $\beta$. We shall analyze below the qualitative behavior of the collapse providing examples for $\beta<\beta_c$ and for $\beta>\beta_c$.

In the presence of an anisotropic pressure component, the null energy condition becomes
\begin{equation}
\label{eq_nec_m_non0}
\frac{\rho+p_n}{\rho_{{\rm max}}} = 4(\tilde{t}^{\,2} - 1) - \frac{\beta^2(\tilde{t}^{\,2} + 1)^{\frac{3}{2}}}{\tilde{r}^{\,3}}>0.
\end{equation}
Note that the anisotropic pressure contributes negatively to this equation, which means that a possible violation of this condition could be faced as a consequence of dissipation and not solely as an exotic fluid contribution. Furthermore, the radial coordinate dependence that appears in this equation can be replaced by the evolution of the star radius indicating the character (trapped or marginal) of $\Sigma$ if the inequality (\ref{eq_nec_m_non0}) is violated.

\begin{figure}
\begin{center}
\includegraphics[width=60mm]{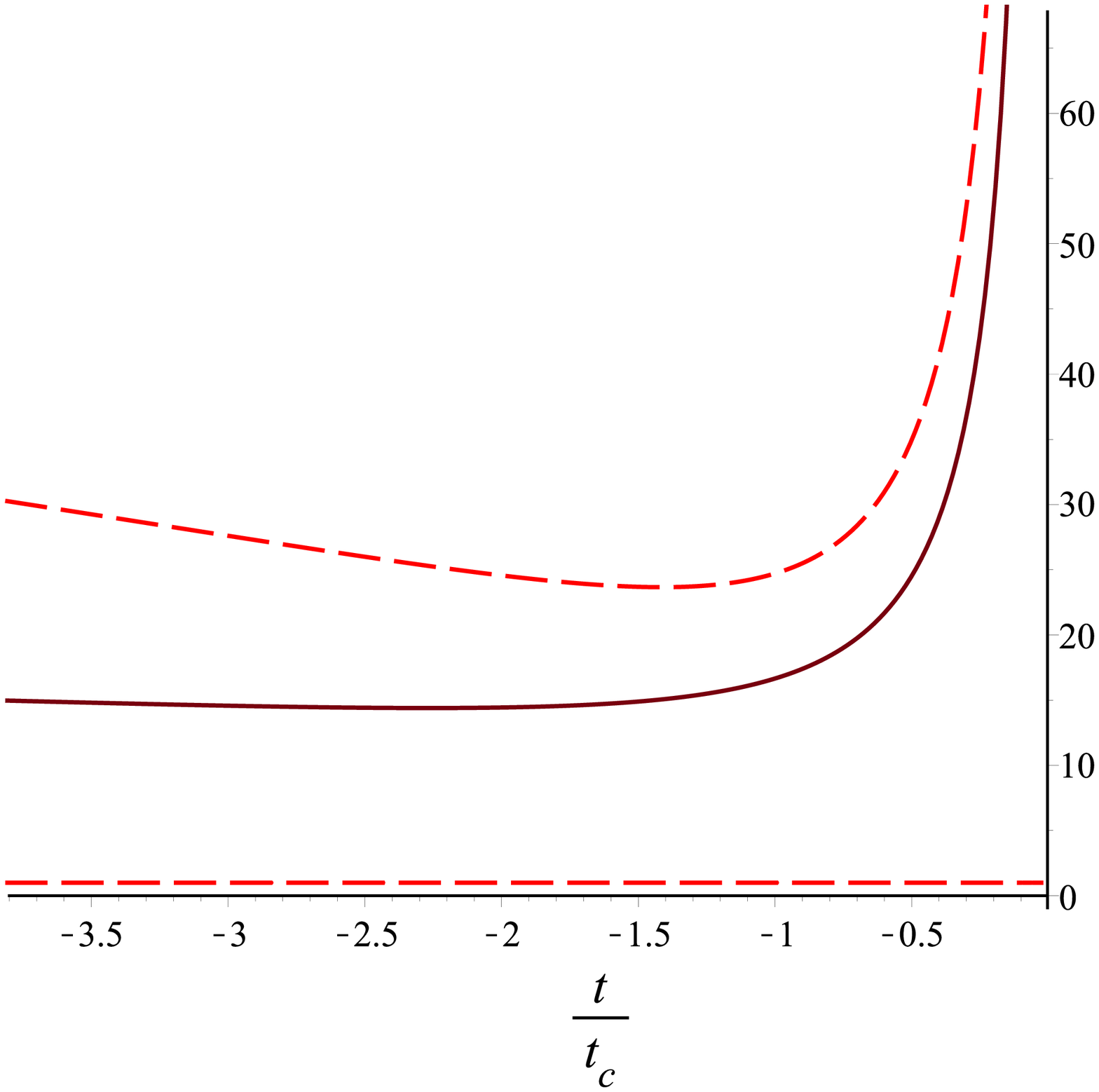}\hspace{.5cm}
\includegraphics[width=60mm]{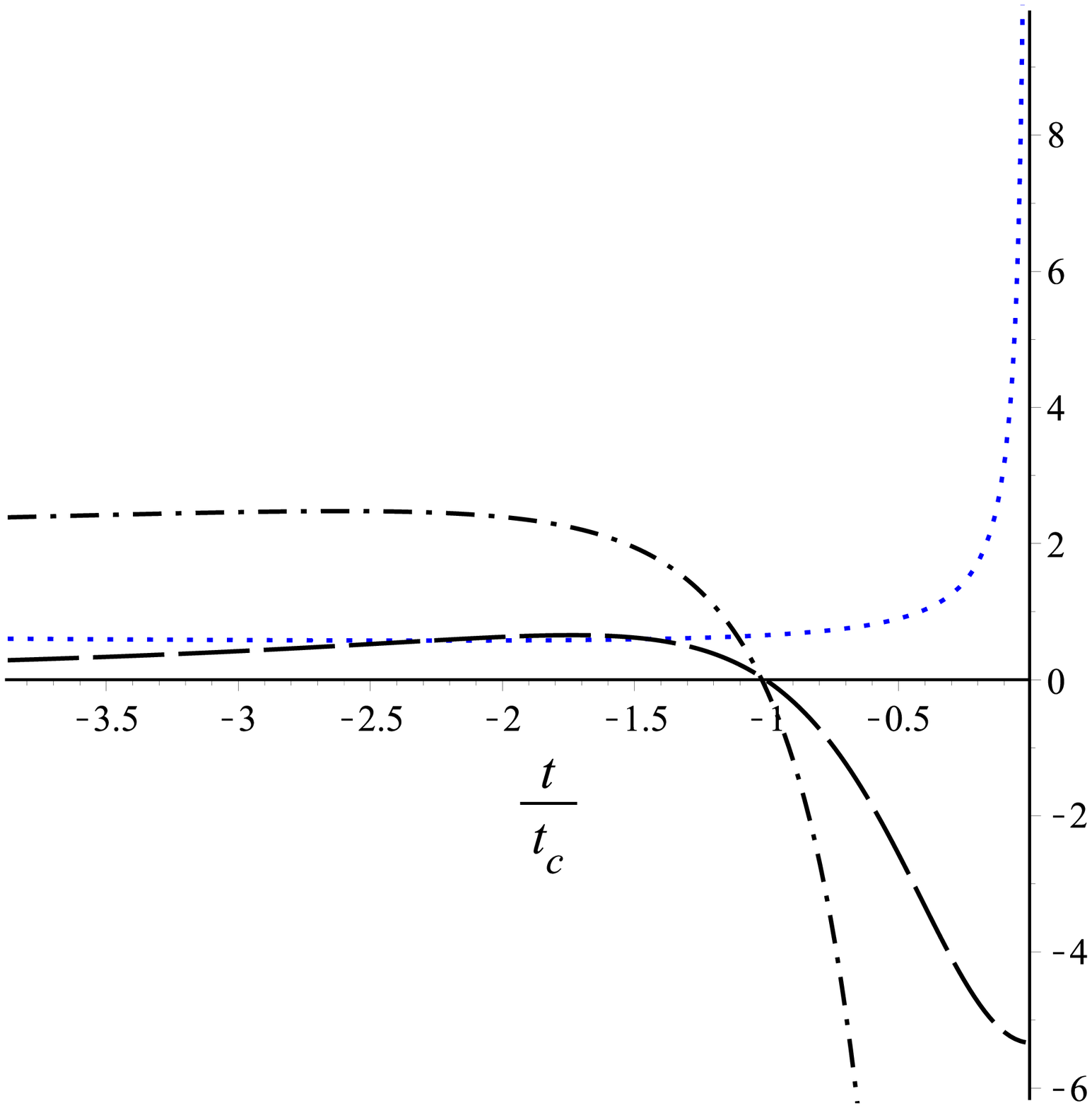}
\end{center}
\caption{Case $M\neq0$. Illustrative initial values are used as $\beta=15$, $\tilde t=-\sqrt{15}$ and $\tilde r(\tilde t)\approx2.45$. On the left panel, it is depicted $\tilde r(\tilde t)$ as solid line and $\tilde r_{A3H}(\tilde t)$ as dashed lines. The region where $\Sigma$ is an untrapped surface is located between the two dashed lines. On the right panel, $d\tilde u/d\tilde t$ (dot-dashed line), the mass function (dotted line) and $\rho+p_n$ (long-dashed line) are presented. The qualitative behavior is similar to Fig.\ (\ref{fig2}).}
\label{fig4}
\end{figure}

In Fig.\ (\ref{fig4}) we depict the behavior of the star radius along time on the left panel. The region between the apparent 3-horizon curves delimits the allowed values of $\tilde{r}(\tilde t)$. With the initial condition used there, it never crosses the horizon, but when the null energy condition is violated [right panel in Fig.\ (\ref{fig4})] it starts growing indefinitely, while the u-coordinate becomes invalid to describe the exterior. This case is comparable with the one depicted in Fig.\ (\ref{fig2}), which is reasonable since one can make the approximation $\beta\gg\beta_c$ implying that $M$ can be neglected and recover most of the results of the previous case. Nevertheless, it is important to recall that the star cannot evaporate when $M\neq0$ even if it is small.

\begin{figure}
\begin{center}
\includegraphics[width=45mm]{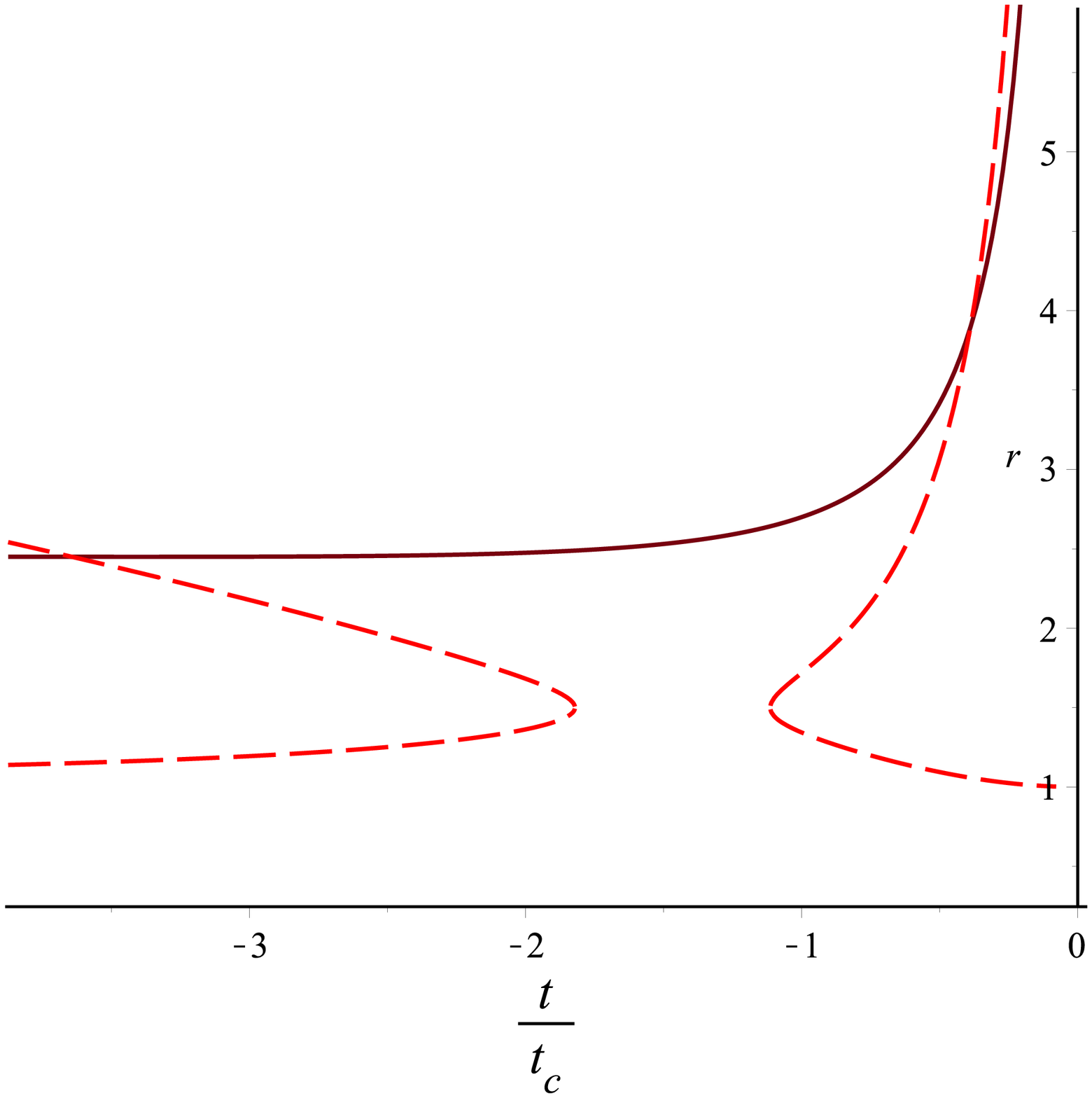}\hspace{.5cm}
\includegraphics[width=45mm]{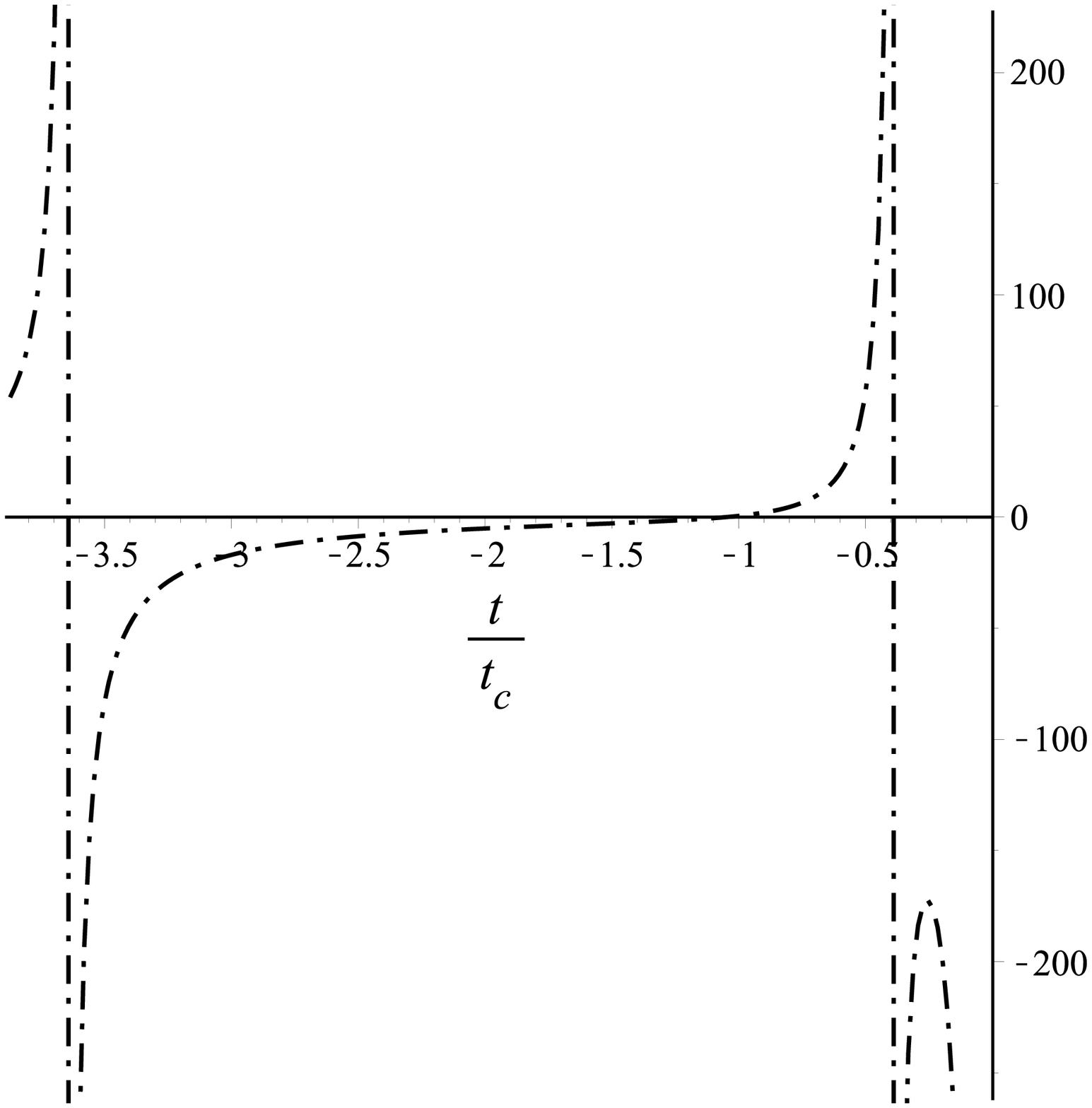}\hspace{.5cm}
\includegraphics[width=45mm]{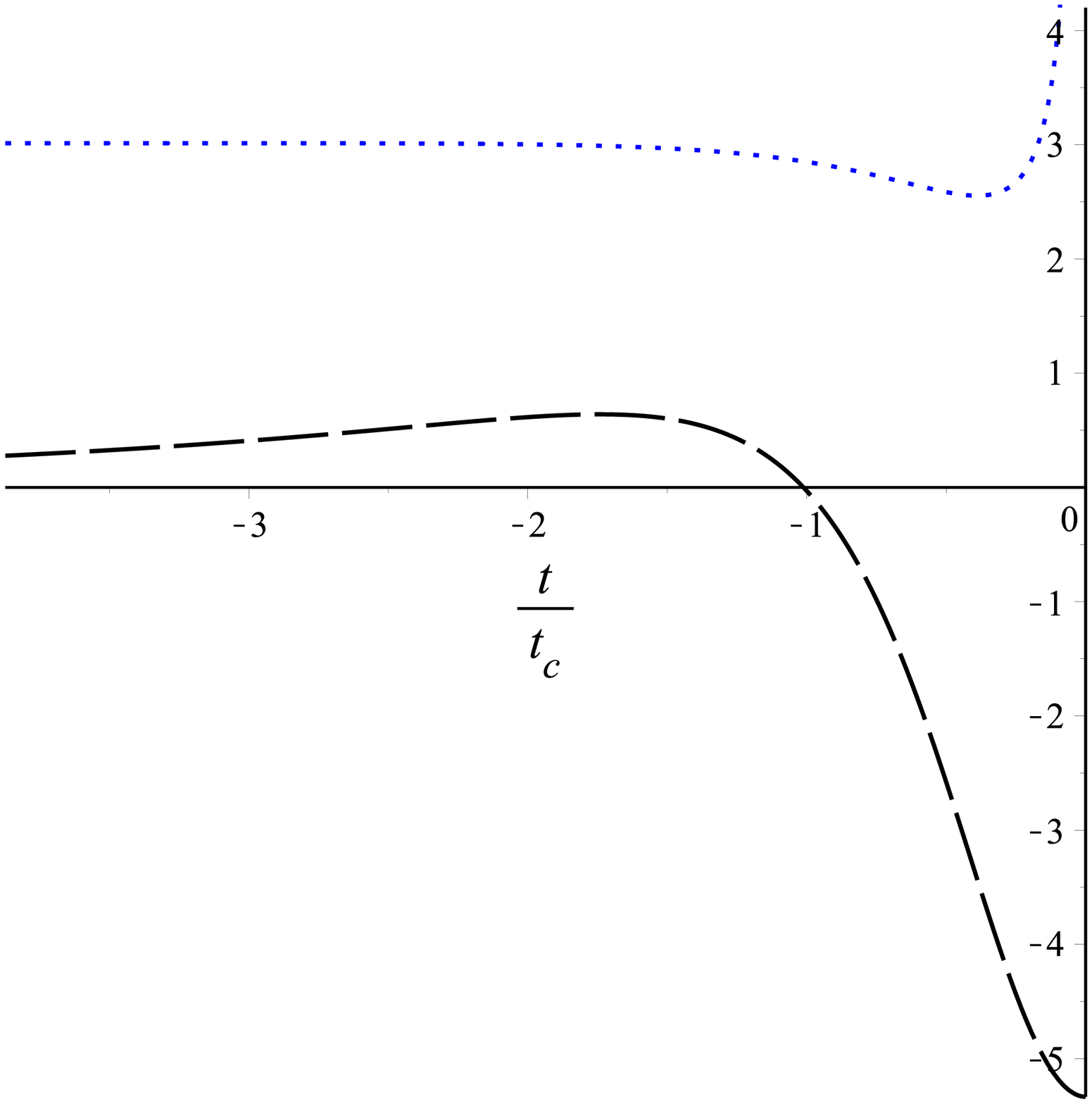}
\end{center}
\caption{Case $M\neq0$. Illustrative initial values are used as $\tilde t=-\sqrt{15}$ and $\tilde r(\tilde t)\approx2.3$. It is depicted $d\tilde u/d\tilde t$ (dot-dashed line), $\tilde r(\tilde t)$ (solid line), $\tilde r_{A3H}(\tilde t)$ (dashed line), the mass function (dotted line) and $\rho+p_n$ (long-dashed line).}
\label{fig5}
\end{figure}

The profile of the apparent 3-horizon is distinct when $\beta<\beta_c$ ($M$ non-negligible). It is possible to show that as $\beta$ approaches its critical value, the dashed lines on the left panel in Fig.\ (\ref{fig4}) become closer, such that they merge in a specific point when $\beta=\beta_c$ and then bifurcate, connecting the regions where $\Sigma$ is a trapped surface and leading to the profile depicted on the left panel of Fig.\ (\ref{fig5}). For the initial conditions we chose ($\tilde{r}_{,\tilde t}<0$ and $\tilde{u}_{,\tilde{t}}>0$) the allowed region becomes more limited in this case, since the left branch of the apparent 3-horizon tends to shrink towards the line $\tilde r=1$ as $\beta$ diminishes. In virtue of this new division of the ($\tilde{r},\tilde{t}$)-plane into regions where $\Sigma$ is trapped, marginal or untrapped, the appearance of an event horizon ($\tilde{u}\rightarrow+\infty$) is unavoidable at some negative value of $\tilde{t}$. Notwithstanding, the interior solution is still regular there, it can be extended beyond this point up to $\tilde t\rightarrow0^{-}$ and all curvature invariants are finite in this limit, similarly to the previous cases.

\subsection{Constant Scale Factor}
For the sake of completeness, we study the special case when the scale factor is constant in time. It means that $a=a_f$, $\rho=\rho_f$ and $p(\rho)=p_f$ are fixed, but $r$ and $\bar m$ can vary with $t$ in principle. Therefore, if $a_{,t}=0$ in (\ref{0}), we obtain:

\begin{equation}
\label{en_den_s_c}
\begin{array}{ll}
\mbox{for}\,\, k=1,&\mbox{then}\,\,  \rho_f=\frac{3}{a_f^2},\\[1ex]
\mbox{for}\,\, k=0,&\mbox{then}\,\,  \rho_f=0, \,\,\,(\mbox{or}\, \rho_f+\Lambda=0, \, \mbox{with}\, \Lambda<0),\\[1ex]
\mbox{for}\,\, k=-1,&\mbox{then}\,\, \rho_f=-\frac{3}{a_f^2}, \,\,\,\left(\mbox{or}\, \rho_f+\Lambda=-\frac{3}{a_f^2}, \, \mbox{with}\, \Lambda<-\frac{3}{a_f^2}\right).
\end{array}
\end{equation}
Note that $k\neq1$ needs a negative cosmological constant in order to have $\rho_f>0$. The substitution of these conditions into (\ref{rdot}) and (\ref{der_mas_func}) leads basically to two different situations: if $k=0$, the equation for $\dot r$ is identically satisfied once both $\rho$ and $p_n$ should be zero. This means that the star radius and its total mass are constant with values given respectively by
\begin{equation}
\label{fin_k0}
r_f=\left(\frac{2M}{p_fa_f^2}\right)^{\frac{1}{3}} \qquad \mbox{and}\qquad \bar m=Ma_f\left(1+\frac{\rho_f}{3p_f}\right).
\end{equation}
Otherwise, if $k\neq0$, then $\dot r$ and $\dot{m}$ still have dynamics, but they have a global attractor given precisely by (\ref{fin_k0}) when the initial conditions are satisfied as described above. From this, we see that even if the scale factor is constant, the gravitational collapse can still occur under limited circumstances. Thus, whenever the first derivative of the scale factor vanishes for an instant of time, the system may undergo a smooth transition to one of the aforementioned cases. In particular, bouncing models are suitable candidates to undergo such transition as $a_{,t}=0$ at the bounce \cite{vanessa}.

\section{Concluding Remarks}\label{VII}

We have analyzed the problem of radiating spherical gravitational collapse by considering an inhomogeneous interior solution where a new parameter $M$ prevents the radial marker of this geometry from vanishing. Although this fact could suggest that a stable object of minimum radius would be produced as an outcome of the collapse, we have shown that such viscous-related quantity actually narrowed the window where suitable initial conditions for a non singular collapse satisfying the energy conditions could be found. This is also consequence of the fact that this parameter forbids the complete evaporation of the star before it crosses the apparent 3-horizon, which was shown in the literature before to result in a regular process where the final configuration is the Minkowski spacetime.

We also scrutinized how the dynamics of the scale factor of the interior geometry affects the collapse, both in the cases $M=0$ and $M\neq 0$, showing that, even when it has a non-singular bouncing behavior, no obvious set of initial conditions exists for a regular collapse. In most of the configurations obtained the star's surface ends up crossing the apparent 3-horizon, becoming a trapped surface which would thus lead to singularities according to the singularity theorems. However, the necessary violation of the energy conditions required by the bounce seems to prevent the singularity formation, smoothing the spacetime curvature and forcing the star to expand infinitely in a finite amount of time as $t$ approaches $0$ from negative values. Certainly, this is not the final statement concerning the gravitational collapse in this case because the maximal analytical extension of the solution is not known, but this may elucidate a few points about the role played by the energy condition violation in gravitational collapse in general.

\ack
The authors are in debt with R. Klippert for his valuable comments on this manuscript. GBS would like to thank the PCI program at the Brazilian Center for Research in Physics--CBPF, where part of this work was developed, for financial support. VPF thanks FAPERJ for the grant E-26/200.279/2015.

\end{document}